\documentclass[12pt,twoside]{article}

\evensidemargin=\oddsidemargin
\def\Title#1#2#3{%
    \baselineskip=18pt
    \begin{center}
          {\Large\bf\uppercase{#1} \\ }
          \bigskip\bigskip
          {#2} \\
          {#3} \\
    \end{center}}

\long\def\Abstract#1{%
         \bigskip
         \parbox{0.93\textwidth}{%
                 \begin{center}
                       {\bf Abstract} \\
                 \end{center}
                 \medskip{\baselineskip=14pt #1}
                 \vss}
         \bigskip}

\begin{document}

\Title{THE STATUS OF THE $\Lambda$ TERM \\
IN QUANTUM GEOMETRODYNAMICS \\
IN EXTENDED PHASE SPACE}%
{T. P. Shestakova}%
{Department of Theoretical and Computing Physics, Rostov State University, \\
Sorge Str. 5, Rostov-on-Don 344090, Russia \\
E-mail: {\tt stp@phys.rnd.runnet.ru}}

\Abstract{S. Weinberg pointed out a way to introduce
a cosmological term by modifying the theory of gravity. This
modification would be justified if the Einstein equations with
the cosmological term could be obtained in the classical limit
of some physically satisfied quantum theory of gravity. We propose
to consider quantum geometrodynamics in extended phase space as
a candidate for such a theory. Quantum geometrodynamics in extended
phase space aims at giving a selfconsistent description of the
integrated system ``the physical object (the Universe) + observation
means'', observation means being represented by a reference frame.
The $\Lambda$ term appears in classical equations under certain gauge
conditions and characterizes the state of gravitational vacuum
related to a chosen reference frame. The eigenvalue spectrum of
$\Lambda$ depends on a concrete cosmological model and can be found
by solving the Schr\"odinger equation for a wave function of the
Universe. The proposed version of quantum geometrodynamics enables one
to make predictions concerning probable values of the $\Lambda$ term
at various stages of cosmological evolution.}

\section{Introduction}
Among various approaches to solving the cosmological constant problem,
S. Weinberg \cite{Weinberg} pointed to the modification of the theory of
gravity when the cosmological term would appear in the Einstein equations
as an integration constant. Weinberg's viewpoint was that this modification
would be justified if the Einstein equations with the cosmological term
could be obtained in the classical limit of some physically satisfied
quantum theory of gravity.

In this paper I shall consider quantum geometrodynamics (QGD) in extended
phase space as a candidate for such a theory. The proposed modification of
QGD enables one to make predictions concerning probable values of the
$\Lambda$ term at various stages of cosmological evolution.

However, this modification is gauge-nonivariant in general. So the question
arises, what grounds do we have for considering a gauge-noninvariant quantum
geometrodynamics?

\section{The grounds for considering a gauge-nonivariant\protect\\
version of quantum geometrodynamics}
Now I shall try to show that we do really have such grounds when
constructing quantum theory of a closed universe. One of the reasons
is that there are no asymptotic states in a closed universe. Meanwhile,
any gauge-invariant quantum field theory is essentially based on {\it the
assumtion about asymptotic states}. In the path integral approach, which is
more adequate for quantizing a gauge theory, asymptotic boundary conditions
ensure gauge invariance of a path integral and plays the role of selection
rules \cite{Hennaux}. However, in a general case without asymptotic states
it is mathematically impracticable to separate physical degrees of freedom
from ``nonphysical'' ones and identify gravitational field with a system
with two field degrees of freedom.

Another argument is the well-known parametrization
noninvariance of the Wheeler -- DeWitt equation (see, for
example \cite{HP,Halliwell}).  However, a transition to another gauge
variable is formally equivalent to imposing a new gauge condition, and vice
versa. The latter reflects an obvious fact that the choice of gauge
variables and the choice of gauge conditions have a unified interpretation:
they together determine equations for the metric components $g_{0\mu}$,
fixing a reference frame.

\begin{center}
\begin{tabular}{ccccc}
Parametrization & + & Gauge conditions & $\Rightarrow$ &
 Equations for $g_{0\mu}$\\
$g_{0\mu}=v_{\mu}\left(\mu_{\nu},\gamma_{ij}\right)$ & &
 $\mu_{\nu}=f_{\nu}\left(\gamma_{ij}\right)$ & &
 $g_{0\mu}=v_{\mu}\left(f_{\nu}\left(\gamma_{ij}\right),\gamma_{ij}\right)$
\end{tabular}
\end{center}

\noindent
Here $\mu_{\nu}$ are new gauge variables, in particular, the lapse and
shift functions, $N$ and $N_i$, $\gamma_{ij}$ is 3-metric.
Thus even if one considers $\mu_{\nu}$ as independent of $\gamma_{ij}$,
different parametrizations will correspond to different reference frames.
So, as a matter of fact, {\it the parametrisation noninvariance of the
Wheeler -- DeWitt equation is ill-hidden gauge noninvariance}.

\section{Quantum geometrodynamics in extended phase space:
a minisuperspace example}
Bearing in mind all the mentioned above, the investigation of a more general
theory seems to be reasonable. In the work by G. M. Vereshkov, V. A.
Savchenko and me \cite{Our1, Our2} quantum geometrodynamics in extended phase
space has been proposed. The extended phase space (EPS) approach developed
by Batalin, Fradkin and Vilkovisky (BFV) \cite{BFV1} -- \cite{BFV4} is
adequate for studying effects related to gauge degrees of freedom.

The central part in our version of QGD is given to the Schr\"odinger equation
for a wave function of the Universe. The Schr\"odinger equation is derived
from a path integral by the standard method \cite{Cheng} originated from
Feynman. In accordance with the physical situation we consider the path
integral without asymptotic boundary conditions. To skeletonize the path
integral the full set of equations in EPS is used. This set of equations is
gauge-noninvariant and nondegenerate. As a result, no ill-definite
mathematical expression arises when deriving the Schr\"odinger equation,
and the procedure of its derivation turns out to be correct but the
Schr\"odinger equation will contain information about parametrization and
gauge.

Let us turn to a minisuperspace model which involves isotropic universe
and Bianchi IX cases.
In a rather broad class of parametrization and gauge conditions the action
in EPS can be reduced to the Faddeev -- Popov effective action:
\begin{equation}
\label{action}
S_{ef\!f}=\!
 \int\!dt\,\biggl\{
  \displaystyle\frac12\frac{v(Q^a)}\mu\gamma_{ab}\dot{Q}^a\dot{Q}^b
  -\frac\mu{v(Q^a)}U\left(Q^a\right)
  +\pi\left(\dot\mu-f_{,a}\dot{Q}^a\right)
  -i\mu\dot{\bar\theta}\dot\theta\biggr\}.
\end{equation}
Here $\mu$ is a new gauge variable defined by
\begin{equation}
\label{mu}
\frac N{a^3}=\frac\mu{v(Q^a)},
\end{equation}
${Q^a}$ are physical variables, $a$ is a scale factor, $\theta,\bar\theta$
are the Faddeev -- Popov ghosts after replacement
$\bar{\theta}\to-i\bar{\theta}$; $\pi$ is a Lagrange multiplier, and
the special class of gauges not depending on time is used
\begin{equation}
\label{mu,f,k}
\mu=f(Q^a)+k;\quad
k={\rm const},
\end{equation}
or, in a differential form,
\begin{equation}
\label{diff_form}
\dot{\mu}=f_{,a}\dot{Q}^a,\quad
f_{,a}\stackrel{def}{=}\frac{\partial f}{\partial Q^a};
\end{equation}
\begin{equation}
\label{3-metric}
\gamma_{ab}=\mathop{\rm diag}\nolimits(-1,\,1,\,1,\,1,\,\ldots).
\end{equation}

The Schr\"odinger equation for this model reads
\begin{equation}
\label{SE1}
i\,\frac{\partial\Psi(Q^a,\mu,\theta,\bar\theta;\,t)}{\partial t}
 =H\Psi(Q^a,\,\mu,\,\theta,\,\bar\theta;\,t),
\end{equation}
where
\begin{equation}
\label{H}
H=\frac i\mu\frac{\partial}{\partial\theta}
  \frac{\partial}{\partial\bar\theta}
 -\frac1{2M}\frac{\partial}{\partial Q^{\alpha}}MG^{\alpha\beta}
  \frac{\partial}{\partial Q^{\beta}}
 +\frac\mu{v(Q^a)}(U+V);
\end{equation}
$M$ is the measure in the path integral,
\begin{equation}
\label{M}
M=v^{\frac K2}(Q^a)\mu^{1-\frac K2};
\end{equation}
\begin{equation}
\label{Galpha_beta}
G^{\alpha\beta}=\frac\mu{v(Q^a)}\left(
 \begin{array}{cc}
  f_{,a}f^{,a}&f^{,a}\\
  f^{,a}&\gamma^{ab}
 \end{array}
 \right);\quad
\alpha,\beta=(0,a);\quad
Q^0=\mu,
\end{equation}
$K$ is a number of physical degrees of freedom; the wave function is defined
on extended configurational space with the coordinates
$Q^a,\,\mu,\,\theta,\,\bar\theta$.
The feature of the present approach is the appearance of a quantum
correction $V$ to the potential $U$. The correction is due to the
dependence of the metric of configurational space of physical
variables, \quad $G^{(phys)}_{ab}=\frac{v(Q^a)}\mu \gamma_{ab}$, \quad
on these variables. $V$ depends on the chosen parametrization and gauge
(\ref{mu}), (\ref{mu,f,k}):
\begin{equation}
\label{V}
V=-\frac{K^2+5K}{24\mu^2}f_{,a}f^{,a}
  +\frac{K+1}{6\mu}f^{,a}_{,a}
  +\frac{K^2-K-2}{12\mu v(Q^a)}v_{,a}f^{,a}
  -\frac{K^2-7K+6}{24v^2(Q^a)}v_{,a}v^{,a}
  +\frac{1-K}{6v(Q^a)}v^{,a}_{,a}.
\end{equation}

The investigation of the set of equations in EPS reveals the existence of
a conserved quantity $E=\mu\dot\pi$.
As a result, the Hamiltonian constraint $H_{ph}=0$ of general relativity is
replaced by the constraint $H=E$, where $H_{ph}$ is a Hamiltonian of
gravitational and matter fields, $H$ is a Hamiltonian in EPS.

The latter means that a Hamiltonian spectrum in the appropriate quantum
theory is not limited by the unique zero eigenvalue. Finding a
spectrum of $E$ becomes one of the main tasks of quantum geometrodynamics in
EPS.

The general solution to the Schr\"odinger equation (\ref{SE1}) has the
following structure:
\begin{equation}
\label{time-depend.WF}
\Psi(Q^a,\,Q^0,\,\theta,\,\bar\theta;\,t)
 =\int\Psi(Q^a)\exp(-iEt)(\bar\theta+i\theta)\,
  \delta(\mu-f(Q^a)-k)\,dE\,dk.
\end{equation}
The dependence of the wave function (\ref{time-depend.WF}) on ghosts is
determined by the demand of norm positivity.
The ``physical part'' of the wave function $\Psi(Q^a)$ is a solution to
the equation
\begin{equation}
\label{phys.SE}
H^0\,\Psi(Q^a)=E\,\Psi(Q^a);
\end{equation}
\begin{equation}
\label{H0}
H^0=\left.\left[-\frac1{2M}\frac{\partial}{\partial Q^a}
  \frac\mu{v(Q^a)}M\gamma^{ab}\frac{\partial}{\partial Q^b}
 +\frac\mu{v(Q^a)}(U+V)\right]\right|_{\mu=f(Q^a)+k}.
\end{equation}

The wave function (\ref{time-depend.WF}) carries the information on
the physical object (the Universe) and a chosen reference frame,
the latter representing the observer in the theory of gravity.
The correlations between the properties of the physical object and those
of the reference frame are manifested in the quantum correction $V$ to the
potential. So quantum geometrodynamics in extended phase space aims at
giving a selfconsistent description of the integrated system ``the physical
object (the Universe) + observation means (a reference frame)''.

The line $E=0$ in the Hamiltonian spectrum is of a particular interest.
If one puts $E=0$ and chooses the gauge $\dot\mu=0$, the Schr\"odinger
equation for the physical part of the wave function will be reduced to the
Wheeler -- DeWitt equation written down for an arbitrary parametrization of
a gauge variable. As we can see, quantum geometrodynamics in EPS involves
the Wheeler -- DeWitt QGD as a particular case. The way of derivation of the
Wheeler -- DeWitt equation demonstrates that the Wheeler -- DeWitt QGD is
not a gauge-invariant theory in a strict mathematical sence. However, taking
a classical limit for the wave function of the state with $E=0$ one obtains
gauge-invariant equations of motion%
\footnote{If the line $E=0$ belongs to a continuous part of the spectrum,
the state with $E=0$ cannot be normalized and therefore is not physical.
To deal with physical states one should consider narrow enough wave packets,
so that the mean value of $E$ with respect to such a packet would be equal
to zero.}.

\section{The cosmological constant problem}
Now let us turn to the cosmological constant problem. Consider
parametrization and gauge
\begin{equation}
\label{Na}
\mu=Na^3;\quad
\mu=1.
\end{equation}
It corresponds to the constraint on the components of 4-metric
\begin{equation}
\label{Wein}
\sqrt{-g}=const.
\end{equation}
As was shown by Weinberg \cite{Weinberg}, imposing the condition
(\ref{Wein}) leads to the appearance of an additional term in the
Einstein equations
\begin{equation}
\label{T_obs}
T^{\nu}_{\mu (obs)}=-\frac1{2\pi^2}\Lambda\delta^{\nu}_{\mu},
\end{equation}
and it follows from the set of equations in EPS that there exist
a conserved quantity
\begin{equation}
\label{E}
E=-\int d^3x\,\sqrt{-g}T^0_{0(obs)}
 =\Lambda.
\end{equation}
The eigenvalue spectrum of $\Lambda$ can be found by solving the
Schr\"odinger equation; results will depend on a chosen cosmological model.

Taking into account the effect of particle creation in the early Universe,
one can consider quantum transisions between states with different values
of $\Lambda$ and make predictions concerning probable values of $\Lambda$ at
various stages of cosmological evolution.

To summarize, in quantum geometrodynamics in extended phase space
the $\Lambda$ term has the following status:
\begin{itemize}
\item The $\Lambda$ term characterizes the state of gravitational vacuum
related to a chosen reference frame.
\item $\Lambda=0$ corresponds to the state of the Universe in which gauge
effects are negligible; the evolution of the Universe can be described in
the classical limit by a gauge-invariant theory;
\item $\Lambda$ may take on nonzero values at an early stage of
cosmological evolution, when gauge-noninvariant effects should be taken into
account.
\end{itemize}

\end{document}